\documentclass[english,10pt,aps,prd,eqsecnum,nofootinbib]{revtex4}
\usepackage{amssymb,amsmath,amsthm,graphicx,amscd}
\usepackage[dvipsnames]{xcolor}
\usepackage{esint,enumerate,comment,ulem}
\usepackage[hidelinks]{hyperref}

\begin{document}

\title{Fluctuation-Dissipation Relation from the \\ Nonequilibrium Dynamics of a Nonlinear Open Quantum System}

\author{Jen-Tsung Hsiang$^{1,2}$}%
\thanks{cosmology@gmail.com}%
\affiliation{$^{1}$ Center for High Energy and High Field Physics, National Central University, Chungli 32001, Taiwan}
\affiliation{$^{2}$ Center for Theoretical Physics, Fudan University, Shanghai 20438, China}

\author{Bei-Lok Hu$^{3}$}%
\thanks{blhu@umd.edu}
\affiliation{$^{3}$Maryland Center for Fundamental Physics and Joint Quantum Institute, \\ University of Maryland, College Park, Maryland 20742, USA}

\begin{abstract}
Continuing our inquiry into the conditions when fluctuation-dissipation relations (FDR) may appear in the context of nonequilibrium dynamics of open quantum systems (over and beyond the conventional FDR from linear response theory) we turn to nonGaussian systems and consider this issue for an anharmonic oscillator interacting with a scalar quantum field bath. We present the general {nonperturbative} expressions for the rate of energy (power) exchange between the anharmonic oscillator and the thermal bath. For the cases that  a stable final equilibrium state exists, and the nonstationary components of the two-point functions of the anharmonic oscillator have negligible contributions to the evaluation of the power balance, we can show nonperturbatively that equilibration implies an FDR for the anharmonic oscillator. We then use a weakly anharmonic oscillator as an example to illustrate that those two assumptions indeed are satisfied according to our first-order perturbative results: that the net energy exchange vanishes after relaxation in the open system dynamics and an equilibrium state exists at late times.
\end{abstract}

\maketitle

\hypersetup{linktoc=all}
\setcounter{tocdepth}{2}

\baselineskip=18pt
\numberwithin{equation}{section}
\numberwithin{figure}{section}

\allowdisplaybreaks

\section{Introduction}

The open system paradigm captures physical reality better than the idealization of a system in total isolation because the environment it interacts often plays a role.  The interlocking relation between the open system and its environment is registered in the fluctuation-dissipation relations (FDRs). While they are rooted in statistical mechanics \cite{Kubo66,KuboBook,Sciama,QTD1,HHAoP} the implications of FDRs are wide ranging, from condensed matter \cite{KadBay,Lovesey}, nuclear/particle (e.g.,\cite{KadBay,FetterWalecka}) to black hole physics (e.g., \cite{CanSci77,Mottola}) and cosmology (e.g., \cite{HuSin95}). Further description of its scope can be found in \cite{Kubo66,GF17,FDRPRD19}. 

In this paper we investigate the FDR for a nonlinear quantum system interacting linearly with a thermal bath: the system of interest is a quantum anharmonic oscillator, the bath is made up of a thermal scalar quantum field. Two primary aspects in statistical mechanics are involved here for both classical and quantum systems: A. FDR in the context of 1) response theory versus 2) nonequilibrium dynamics; B. FDR for 1) linear vs 2) nonlinear systems. The case of FDR in linear response theory (LRT) -- linear systems treated in response theory (A1+B1) is well known from standard textbooks and discussed most widely. For nonlinear systems treated by response theory (A1+B2) both in the near-equilibrium and non-equilibrium contexts,  there is also a long history of significant theoretical development by authors like Kadanoff and Baym \cite{KadBay}, Eremov \cite{Fremov}, Golden et al \cite{Golden}, Zwangzig \cite{ZwaNlL}, Langreth \cite{Langreth}, Zhou et al. \cite{ZhouCD}, Bochkov and  Kuzovlev \cite{BocKuz12} and Stratonovich \cite{Stran}. 

Before we mention some more recent developments,  for clarification purpose,  it is perhaps useful to highlight the differences (feature A above) in the formulation of FDR between the nonequilibrium dynamics (NEq) approach which we follow in our work and the conventional linear (LRT) or nonlinear response theory . 

\paragraph{FDR in LRT vs NEq}

The main differences  in the set-up, the conditions and the features have been described in Sec. 3 of \cite{QRad19}. We summarize them as follows:   

i) \textit{Set-up}:  FDR under LRT operates under the following assumptions: the system of interest (i) has been  prepared in a thermal state in thermal equilibrium with the bath, and  {then} (ii) is taken away  by a weak external disturbance and the corresponding responses  is recorded. By comparison, in the nonequilibrium (NEq) formalism, the system can  start in any state, which may be very different from the thermal state at the bath temperature or the equilibrium state the system finally settles in. Once the initial state of the system and the properties of the bath are given, their interaction determines the evolution history described by the reduced dynamics of the system. If the dynamics of the system comes to equilibration, then an FDR will be determined for that equilibrium state.

ii) \textit{Conditions}: In LRT the equilibrium state of the system which is assumed to hold for all times.  The FDR in LRT is formulated {with respect to this equilibrium state} in terms of perturbation theory -- weak coupling with the bath and small {perturbation} from the equilibrium state. In nonlinear response, nonlinear perturbance is considered, but response functions are well defined and calculated assuming that the perturbance does not exceed a certain limit. In NEq context, the system and its environment can be strongly coupled while dynamically evolving{, but the existence of an equilibrium state is not a priori known}. One needs to first examine if the system will relax to an equilibrium state before one attempts to verify that an FDR exists. Thus FDR in NEq context is an emergent phenomenon depending on many factors which enter into determining the dynamics of the open system.  

iii) \textit{Features}: In the NEq dynamics approach the  equilibrium state which an open system evolves to {in general} is not a Gibbs (thermal) state. The temperature is ill-defined in a dynamical setting only if/until the system reaches equilibrium, whence it can be identified from the reduced density matrix {or the physical observables} of the equilibrated system. Even so it is at best an ``effective'' temperature because it depends on the details of the system and the bath parameters. The exceptional situation is when the  coupling strength between the system and the bath is vanishingly small, which is a preamble of LRT. {Thus in the context of NEq formulation, the temperature appearing in the proportionality constant of the FDR is in fact the initial temperature of the bath, not the temperature of the system in the final equilibrium state, in contrast to LRT.}

In essence, the FDR in LRT plays a spectator role, connecting the response of the system to an external disturbance. In comparison, in the NEq formulation, FDR has a dynamical significance since it ensures the balance of the energy flow between the reduced system and the environment~\cite{QRad19}{, which in turn signifies dynamical relaxation of the system into equilibrium}.

\paragraph{Nonequlibrium dynamics and dynamical response of nonlinear systems}

An important class of problems where it is desirable to follow the   nonequilibrium dynamics of an open system is quantum transport. There is a vast literature on this subject. Suffice it for our purpose here to mention the reviews of Dhar \cite{Dhar} and  Li et al. \cite{LiRen}. As for methodologies closest to ours, we mention the density matrix approach \cite{DSH}, the nonequilibrium Green function approach~\cite{WALT} and most relevantly, the closed time path (CTP, Schwinger-Keldysh, or `in-in') formalism~\cite{ctp} which has been utilized by  e.g., Wang and Heniz~\cite{WH02}, to derive a nonlinear generalization of the fluctuation-dissipation theorem (for the $n$-point Green functions and the amputated one particle irreducible vertex functions) at finite temperature. (See also the work of Carrington et al \cite{Carrington}.  The methods used in the work of Miyazaki and Reichman \cite{MiyRei}, Maciejko et al. \cite{MacWanGuo}, Motz et al. \cite{Motz} are also of interest.  As for the nonequilibrium dynamics of a quantum anharmonic oscillator open system, the literature is also quite vast. For quantum thermal transport, we mention the work of He et al.~\cite{HT16}, where the authors find an effective harmonic theory substitute by applying the self-consistent phonon theory. For quantum chaos (which by design is outside of the parameter range of our treatment) see, e.g., earlier work of Habib's group \cite{Habib} and more recent work of Ueda's group \cite{Ueda}.

\paragraph{This work: FDR from the nonequilibrium dynamics of non-Gaussian systems}

In our recent work we have explored the FDR in the NEq context in two settings: A) coupled harmonic oscillators sharing a common thermal scalar field bath and FDR at late times~\cite{QTD1,FDRPRD19}, B) a chain of   harmonic  oscillators, with endings connected to their own baths, and FDR in a nonequilibrium steady state~\cite{HHAoP}.  Both cases are Gaussian in the system-bath set up which enables us to produce exact formal solutions for the dynamics.  In this work we {take a baby step toward nonGaussian systems\footnote{We note that nonGaussian systems are wide ranged and their behavior can be very different, e.g., mixing and ergodicity can be violated and there is no guarantee of an FDR \cite{nGtransp}.}, in treating nonlinear quantum systems with the help of  perturbative methods.} We consider the existence of FDR for an anharmonic oscillator interacting with a scalar quantum field bath. We use the functional methods originated in \cite{HPZ93} and developed for this problem in \cite{NENL1} to formulate a general,  {nonperturbative} expressions for the rate of energy (power) exchange between the anharmonic oscillator and the thermal bath.  {Under the assumptions} that a stable unique final equilibrium state exists, and the nonstationary components of the two-point functions of the anharmonic oscillator have negligible contributions to the evaluation of the power balance, we can show nonperturbatively that equilibration implies an FDR for the anharmonic oscillator.   The nontrivial $n$-point functions, with $n>2$, of this nonGaussian system do not  play any {explicit} role in the derivation of the FDR. We then use a weakly anharmonic oscillator as an example to illustrate that those two assumptions indeed are satisfied according to our first-order perturbative results: that the net energy exchange vanishes after relaxation in the open system dynamics, and an equilibrium state exists at late times.

This paper is organized as follows. In Sec~\ref{S:drnkw}, we briefly summarize the essence  of the functional method adopted here for the problem of a quantum anharmonic oscillator coupled to a thermal field, and highlight the results~\cite{NENL1} regarding the late-time behaviors of the Green's functions, in particular, the retarded Green's function and the Hadamard function, of the oscillator to the first order in the anharmonic potential of the form $\frac{\lambda}{4!}\,\chi^{4}$ with $\lambda>0$. In Sec.~\ref{S:fieryga}, we formally derive a nonperturbative expression for the net energy exchange between the anharmonic oscillator and the bath field, and argue that based on the prerequisite assumptions, the balance of this energy flow can imply a nonperturbative  FDR for the anharmonic oscillator after its dynamics is relaxed to the final equilibrium state. Finally, in Sec.~\ref{S:ierbdsfwea}, we apply the functional perturbative approach to a weakly anharmonic oscillator, and show that to the first order in the anharmonic potential, the energy flow between the oscillator and the field bath does come to balance without preconditions, and derive an FDR for the first-order corrections of the anharmonic oscillator's noise and dissipation kernels, that is, a special case of the nonperturbative FDR discussed in Sec.~\ref{S:fieryga}.


\section{Nonequilibrium Evolution of a Driven Anharmonic Oscillator in a Thermal Bath}\label{S:drnkw}
\subsection{in-in generating functional}

The action for an anharmonic oscillator of mass $m$ and  bare natural frequency $\omega>0$  coupled with arbitrary strength $e$ to a bath of massless quantum scalar field $\phi(x)$, initially (at $t=0$) prepared in a thermal state, and driven by an external current $j$, is given by
\begin{align}\label{E:rdher}
	 S_{V}[\chi,\phi]&=\int_{0}^{t}\!ds\;\Bigl\{\frac{m}{2}\Bigl[\dot{\chi}^{2}(s)-\omega^{2}\chi^{2}(s)\Bigr]-V[\chi(s)]+j(s)\chi(s)\Bigr\}\\
	 &\qquad\qquad\qquad\qquad+\int^{t}_{0}\!d^{4}x\;e\chi(s)\delta^{3}(\mathbf{x}-\mathbf{z}(s))\phi(x)+\int^{t}_{0}\!d^{4}x\;\frac{1}{2}\,\bigl[\partial_{\mu}\phi(x)\bigr]\bigl[\partial^{\mu}\phi(x)\bigr]\Bigr\}\,,\notag
\end{align}
The anharmonic potential $V$ will be chosen to be a monomial in the displacement $\chi$ of the oscillator, although the functional formalism we adopt here is not restricted to this condition. The self-coupling constant $\lambda$ associated with the potential will be assumed to be sufficiently weak. The parameter $\mathbf{z}$ denotes the location of the oscillator. Thus Eq.~\eqref{E:rdher}  describes the case of an anharmonic oscillator in the dipole approximation, also known as an Unruh-DeWitt detector,  whose internal degree of freedom is now modeled by an \textsl{an}harmonic oscillator.

In \cite{NENL1}, we have shown that given the initial state of the oscillator at $t=0$, the reduced density matrix $\rho_{\chi}^{(V)}$ of the anharmonic oscillator at a later time $t$ is given by
\begin{align}
	 \rho_{\chi}^{(V)}(q_{b},r_{b},t)&=\exp\biggl\{-i\int_{0}^{t}\!ds\;\Bigl[V(\frac{\delta}{i\,\delta j_{+}})-V(-\frac{\delta}{i\,\delta j_{-}})\Bigr]\biggr\}\,\rho_{\chi}(q_{b},r_{b},t)\;\bigg|_{j_{q}=0=j_{r}}\,,
\end{align}
where $\rho_{\chi}(q_{b},r_{b},t)$ is the reduced density matrix element of the free oscillator, and
\begin{align}
	q&=\frac{\chi_{+}+\chi_{-}}{2}\,,&\frac{\delta}{\delta j_{+}}&=+\frac{\delta}{\delta j_{q}}+\frac{1}{2}\frac{\delta}{\delta j_{r}}\,,\\
	r&=\chi_{+}-\chi_{-}\,,&\frac{\delta}{\delta j_{-}}&=-\frac{\delta}{\delta j_{q}}+\frac{1}{2}\frac{\delta}{\delta j_{r}}\,.
\end{align}
Thus the expectation value of an operator $\mathcal{O}$ consisting only of the anharmonic oscillator variables is given by
\begin{align}\label{E:dkjeir}
	 \langle\,\mathcal{O}(t)\,\rangle&=\frac{1}{\mathcal{Z}_{V}}\,\operatorname{Tr}\Bigl\{\mathcal{O}\,\rho^{(V)}_{\chi}(t)\Bigr\}=\frac{1}{\mathcal{Z}_{V}}\exp\biggl\{-i\int_{0}^{t}\!ds\;\Bigl[V(\frac{\delta}{i\,\delta j_{+}})-V(-\frac{\delta}{i\,\delta j_{-}})\Bigr]\biggr\}\langle\,\mathcal{O}\,\rangle_{0}\,\mathcal{Z}_{}\;\bigg|_{j_{q}=0=j_{r}}\,,
\end{align}
with $\mathcal{Z}=\operatorname{Tr}\bigl\{\rho_{\chi}(t)\bigr\}$, $\mathcal{Z}_{V}=\operatorname{Tr}\bigl\{\rho_{\chi}^{(V)}(t)\bigr\}$ being the `in-in' generating functional of the free oscillator and the anharmonic oscillator respectively at the final time $t$ to ensure proper normalization. The subscript $0$ on an expectation value $\langle\,\mathcal{O}\,\rangle_{0}$ refers  to the results without the anharmonic potential.

The generating functional $\mathcal{Z}$ has been shown~\cite{NENL1} to take on the form
\begin{align}\label{E:ngmshtse}  
	 \mathcal{Z}[j_{q},j_{r}]&=\exp\biggl\{-\frac{1}{4}\int_{0}^{t}\!ds\!\int_{0}^{t}\!ds'\;j_{q}(s)\Bigl[\sigma^{2}D_{1}(s)D_{1}(s')+\frac{1}{m^{2}\sigma^{2}}\,D_{2}(s)D_{2}(s')\Bigr]j_{q}(s')\biggr.\\
	 &\qquad\qquad\quad+\frac{i}{m}\int_{0}^{t}\!ds\!\int_{0}^{s}\!ds'\;j_{q}(s)\,D_{2}(s-s')\,j_{r}(s')-\biggl.\frac{e^{2}}{2}\int_{0}^{t}\!ds\!\int_{0}^{t}\!ds'\;\mathfrak{J}_{q}(s)\,G_{H,\,\beta}^{(\phi)}(s-s')\,\mathfrak{J}_{q}(s')\biggr\}\,,\notag
\end{align}
when the initial state of the oscillator is a wavepacket of width $\sigma$
\begin{equation}\label{E:eubkjfg}
	 \rho_{\chi}(q_{a},r_{a},0)=\left(\frac{1}{\pi\sigma^{2}}\right)^{1/2}\exp\biggl\{-\frac{1}{\sigma^{2}}\left[r_{a}^{2}+\frac{1}{4}\,q_{a}^{2}\right]\biggr\}\,.
\end{equation}
$\mathfrak{J}_{q}(s)$ is a shorthand notation for the integral
\begin{equation}\label{E:rktdhs}
	\mathfrak{J}_{q}(s)=\frac{1}{m}\int_{0}^{t}\!ds'\;D_{2}(s'-s)\,j_{q}(s')\,,
\end{equation}
and $D_{1,2}(t)$ are defined in \eqref{E:dbieusfs}. The full generating functional $\mathcal{Z}_{V}$ in the presence of a nonlinear potential $V$ {in polynomial form} can be expanded by
\begin{align}
	\mathcal{Z}_{V}&=\exp\biggl\{-i\int_{0}^{t}\!ds\;\Bigl[V(\frac{\delta}{i\,\delta j_{+}})-V(-\frac{\delta}{i\,\delta j_{-}})\Bigr]\biggr\}\,\mathcal{Z}=\mathcal{Z}+\mathcal{Z}_{1}+\cdots\,,\label{E:dvjgrtd}
\end{align}
where $\mathcal{Z}_{1}$ is the leading order correction of $\mathcal{Z}$ due to the nonlinear potential $V$, {assuming the nonlinearity is relatively weak}. It has been shown~\cite{NENL1} that in the limit $j\to0$, there is no first-order correction of $\mathcal{Z}_{V}$ due to the nonlinear potential, and then
\begin{equation}\label{E:dgetys}
	\mathcal{Z}_{V}=\mathcal{Z}_{}+\mathcal{O}(\lambda^{2})\,.
\end{equation}

Eq.~\eqref{E:dkjeir} already supplies us the information about the nonequilibrium evolution of the anharmonic oscillator when it is coupled to a thermal bath. For the purpose of this paper  the dynamics of the real-time two-point functions of the anharmonic oscillator is of special interest.

\subsection{real-time two-point functions}

We now give a brief derivation via the functional method of the first-order correction to the two-point functions of the anharmonic oscillator. Further details can be found in~\cite{NENL1}. For a quartic anharmonic potential
\begin{equation}\label{E:kdjvhgre}
	V(\chi)=\frac{\lambda}{4!}\,\chi^{4}\,,\qquad\qquad\qquad\text{with $\lambda>0$}\,,
\end{equation}
the  first-order correction of the generating functional is  given by
\begin{align}
	\mathcal{Z}_{1}[j]&=-i\lambda\int_{0}^{t}\!ds\;\biggl\{\frac{1}{2!}\,\mathfrak{J}_{q}(s)C(s,s)\Xi(s)\mathcal{Z}+\frac{1}{3!}\,\mathfrak{J}_{q}(s)\Xi^{3}(s)\mathcal{Z}+\frac{1}{4!}\,\mathfrak{J}_{q}^{3}(s)\Xi(s)\mathcal{Z}\biggr\}\,.\label{E:hgvjrts}
\end{align}
after carrying out the functional derivatives according to \eqref{E:dvjgrtd} and \eqref{E:kdjvhgre}, where $\mathfrak{J}_{q}$ is given by \eqref{E:rktdhs}, and 
\begin{align*}
	\Xi[j;\tau)&=\frac{i}{2}\int_{0}^{t}\!ds'\;\Bigl[\sigma^{2}D_{1}(\tau)D_{1}(s')+\frac{1}{m^{2}\sigma^{2}}\,D_{2}(\tau)D_{2}(s')\Bigr]j_{q}(s')+\frac{1}{m}\int_{0}^{t}\!ds'\;D_{2}(\tau-s')\,j_{r}(s')\notag\\
	&\qquad\qquad\qquad\qquad\qquad\qquad\qquad+i\,\frac{e^{2}}{m}\int_{0}^{t}\!ds\!\int_{0}^{t}\!ds'\;D_{2}(\tau-s)\,G_{H,\,\beta}^{(\phi)}(s-s')\,\mathfrak{J}_{q}(s')\,,\\
	C(\tau,\tau')&=D_{1}(\tau)D_{1}(\tau')\,\langle\hat{\chi}^{2}(0)\rangle+D_{2}(\tau)D_{2}(\tau')\,\langle\dot{\hat{\chi}}^{2}(0)\rangle\\
	&\qquad\qquad\qquad\qquad\qquad\qquad\qquad+\frac{e^{2}}{m^{2}}\int_{0}^{\tau}\!ds\int_{0}^{\tau'}\!ds'\;D_{2}(\tau-s)D_{2}(\tau'-s')G_{H,0}^{(\phi)}(\mathbf{z},s;\mathbf{z},s')\,.
\end{align*}
The real-time Green's functions of the anharmonic oscillator can be constructed from the path-ordered two-point functions
\begin{align}
	\langle\,\mathcal{P}\chi(\tau)\chi(\tau')\,\rangle=\begin{cases}
		\langle\,\mathcal{T}\chi(\tau)\chi(\tau')\,\rangle\,,&\tau\in C_{+}\;\&\;\tau'\in C_{+}\,,\\
		\langle\,\chi(\tau')\chi(\tau)\,\rangle\,,&\tau\in C_{+}\;\&\;\tau'\in C_{-}\,,\\
		\langle\,\chi(\tau)\chi(\tau')\,\rangle\,,&\tau\in C_{-}\;\&\;\tau'\in C_{+}\,,\\
		\langle\,\mathcal{T}^{*}\chi(\tau)\chi(\tau')\,\rangle\,,&\tau\in C_{-}\;\&\;\tau'\in C_{-}\,,
	\end{cases}
\end{align}
where $C_{+/-}$ represents the forward/backward time branch and $\mathcal{T}$, $\mathcal{T}^{*}$ denote time-ordering and anti-time-ordering. Thus, the Feynman propagator of the anharmonic oscillator can be evaluated as the second derivatives of the generating functional with respect to $j_{+}$ at two different times, that is, with $0<\tau,\,\tau'<t$. Its first-order correction is then  
\begin{align}
	&\quad\langle\,\mathcal{T}\,\chi(\tau)\chi(\tau')\,\rangle^{(1)}=-\frac{1}{\mathcal{Z}[j;t)}\frac{\delta^{\,2}\mathcal{Z}_{1}[j;t)}{\delta j_{+}(\tau)\,\delta j_{+}(\tau')}\;\bigg|_{\substack{j=0\\q=0}}\,.
\end{align}
which, with the help of \eqref{E:hgvjrts}, becomes
\begin{align}
	\langle\mathcal{T}\chi(\tau)\chi(\tau')\rangle^{(1)}&=\lambda\int_{0}^{t}\!ds\,\biggl\{-\frac{1}{2m}\,D_{2}(\tau-s)C(s,s)C(s,\tau')-\frac{1}{2m}\,D_{2}(\tau'-s)C(s,s)C(s,\tau)\biggr.\\
	&\qquad\qquad\qquad\qquad+\biggl.\frac{i}{4m^{2}}\biggl[D_{2}(\tau-s)C(s,s)D_{2}(s-\tau')+D_{2}(\tau'-s)C(s,s)D_{2}(s-\tau)\biggr]\biggr\}\,,\notag
\end{align}
This allows us to read off~\cite{NENL1} the first-order corrections of the Hadamard function and the retarded Green's function of the anharmonic oscillator,
\begin{align}
	G_{H,1}^{(\chi)}(\tau,\tau')&=-\frac{\lambda}{2}\int_{0}^{t}\!ds\;\Bigl[G_{R,0}^{(\chi)}(\tau-s)G_{H,0}^{(\chi)}(s,s)G_{H,0}^{(\chi)}(s,\tau')+G_{R,0}^{(\chi)}(\tau'-s)G_{H,0}^{(\chi)}(s,s)G_{H,0}^{(\chi)}(s,\tau)\Bigr]\,,\label{E:oerwpijen1}\\
	G_{R,1}^{(\chi)}(\tau,\tau')&=-\frac{\lambda}{2}\int_{0}^{t}\!ds\;G_{R,0}^{(\chi)}(\tau-s)G_{H,0}^{(\chi)}(s,s)G_{R,0}^{(\chi)}(s-\tau')\,,\label{E:oerwpijen2}
\end{align}
with $0<\tau'\leq\tau<t$. The Green's functions~\cite{NENL1}
\begin{align}
	G_{R,0}^{(\chi)}(\tau-\tau')&=\frac{1}{m}\,D_{2}(\tau-\tau')\,,&G_{H,0}^{(\chi)}(\tau,\tau')&=C(\tau,\tau')\,.
\end{align}
are the zeroth-order Green's functions of the anharmonic oscillator, that is, the Green's functions of the harmonic oscillator coupled to the scalar field. Eqs.~\eqref{E:oerwpijen1} and \eqref{E:oerwpijen2} imply that these first-order corrections in general are not stationary.

However, it can be shown~\cite{NENL1} that the first-order corrections of the retarded Green's function and the Hadamard function of the anaharmonic oscillator will become stationary at late times, as their zeroth-order counterparts do. We then have
\begin{align}\label{E:dkgdnksd}
	G_{H,1}^{(\chi)}(\tau,\tau')&=G_{H,1}^{(\chi)}(\tau-\tau')\,,&G_{R,1}^{(\chi)}(\tau,\tau')&=G_{R,1}^{(\chi)}(\tau-\tau')\,,
\end{align}
for $\gamma^{-1}\ll\tau,\,\tau'$, where $\gamma=e^{2}/8\pi m$ is the damping constant. This nice property can be partly traced back to the consequence of the interaction between the oscillator and the quantum field. In the context of the perturbative treatment, we observe that the zeroth-order dynamics of the oscillator is equivalent to a damped harmonic oscillator, driven by a stochastic force, or noise, representing the quantum fluctuations of the field and inherits its statistical {properties, such as spectral density, etc. This noise from the environment imparts a stochastic component in the motion of the oscillator which generates radiation whose backreaction introduces a dissipative force which   dampens the oscillator's motion. (See, e.g., \cite{QRad19} for a fuller description.}  So long as the displacement of the oscillator and the noise force are sufficiently small, it is plausible that the anharmonic oscillator can still relax to a steady state. If the displacement (caused by the driving force) is not small enough, then the anharmonic potential may excite the oscillator to a higher energy, counteracting the energy loss due to dissipation. Under such circumstances the dynamics of the nonlinear oscillator can become rather complicated, and the perturbation expansion to the first order may cease to be valid.

The stationarity of the late-time dynamics of the anharmonic oscillator implies that a  fluctuation-dissipation relation exists at the first-order correction of the anharmonic oscillator's Green's functions
\begin{equation}\label{E:rturtdfjw}
	\widetilde{G}_{H,1}^{(\chi)}(\kappa)=\coth\frac{\beta\kappa}{2}\,\operatorname{Im}\widetilde{G}_{R,1}^{(\chi)}(\kappa)\,,
\end{equation}
at late time, in additional to the zeroth-order (linear or harmonic oscillator) counterparts
\begin{equation}\label{E:rturtdfjw2}
	\widetilde{G}_{H,0}^{(\chi)}(\kappa)=\coth\frac{\beta\kappa}{2}\,\operatorname{Im}\widetilde{G}_{R,0}^{(\chi)}(\kappa)\,,
\end{equation}
as well as the corresponding relation for the free scalar field
\begin{equation}\label{E:rturtdfjw3}
	\widetilde{G}_{H,\beta}^{(\phi)}(\kappa)=\coth\frac{\beta\kappa}{2}\,\operatorname{Im}\widetilde{G}_{R,0}^{(\phi)}(\kappa)\,.
\end{equation}
It is interesting to note~\cite{QRad19} that they all have the same proportionality factor $\displaystyle\coth\frac{\beta\kappa}{2}$, which depends on the initial temperature of the scalar field, not the `temperature' of the oscillator. This seems to be a generic feature of the nonequilibrium dynamical descriptions of quantum open systems. The oscillator will also inherit this temperature only if the coupling between the oscillator and the bath field is vanishingly weak. Otherwise, the effective temperature of the oscillator will depend~\cite{QTD1,MU19} on the configuration. Eqs.~\eqref{E:rturtdfjw} and \eqref{E:rturtdfjw2} seem to light up the hope that we may still obtain a similar form for the higher-order corrections of the Green's function of the anharmonic oscillator, as long as the perturbative expansion remains valid for all times. We will show in the next section that under certain assumptions, we can indeed give a nonperturbative derivation of the FDR at late times for an anharmonic oscillator coupled to a quantum field bath. Moreover, we can provide a derivation of \eqref{E:rturtdfjw} from a  more physically transparent perspective.

In fact, the derivation of \eqref{E:rturtdfjw} has implied two important conditions: 1) The zeroth-order dynamics has a steady state at late times. This state is approached at exponential time and behaves like an attractor, independent of the initial conditions of the oscillator. 2) The nonstationary components of the two-point functions of a weakly nonlinear oscillator vanish exponentially fast at late times. This condition is  related to the first condition regarding the exponential relaxation of the dynamics. We will see that these two conditions are essential to providing the nonperturbative arguments for the anharmonic oscillator in next section.

\section{Energy Flow Balance between an Anharmonic Oscillator and its Quantum Field Bath - Nonperturbative Arguments}\label{S:fieryga}

From the simultaneous set of Heisenberg equations under consideration
\begin{align}
	\ddot{\hat{\chi}}(t)+\omega^{2}\hat{\chi}(t)+\lambda\,V'[\hat{\chi}(t)]&=\frac{e}{m}\,\hat{\phi}(\mathbf{z},t)\,,\label{E:vher1}\\
	\bigl(\partial_{t}^{2}-\pmb{\nabla}_{x}^{2}\bigr)\hat{\phi}(\mathbf{x},t)&=e\,\hat{\chi}(t)\,\delta^{(3)}(\mathbf{x}-\mathbf{z})\,,\label{E:vher2}
\end{align}
we have the solution of \eqref{E:vher2} given by
\begin{equation}\label{E:vher3}
	\hat{\phi}(\mathbf{x},t)=\hat{\phi}_{h}(\mathbf{x},t)+e\int\!d^{4}x'\;G_{R,\,0}^{(\phi)}(\mathbf{x},t;\mathbf{x}',s)\,\hat{\chi}(s)\,\delta^{(3)}(\mathbf{x}'-\mathbf{z})\,.
\end{equation}
Substituting \eqref{E:vher3} into \eqref{E:vher1} gives the reduced Heisenberg equation for the nonlinear oscillator
\begin{align}\label{E:djsh}
	\ddot{\hat{\chi}}(t)+\omega^{2}\hat{\chi}(t)+\lambda\,V'[\hat{\chi}(t)]-\frac{e^{2}}{m}\int_{0}^{t}\!ds\;G_{R,\,0}^{(\phi)}(\mathbf{z},t;\mathbf{z},s)\,\hat{\chi}(s)&=\frac{e}{m}\,\hat{\phi}_{h}(\mathbf{z},t)\,.
\end{align}
As we have argued before, the nonlinear potential $V(\hat{\chi})$ must possess certain nice features to possibly ensure a stable and unique final state of the oscillator. The term $e\hat{\phi}_{h}(\mathbf{x},t)$ on the righthand side of \eqref{E:djsh} represents the noise force from the field bath. The second term on the other hand accounts for the backaction from the radiation of the scalar field $\phi$ emitted from the nonlinear oscillator. These two terms, originating from the interaction between the oscillator and the field, will serve as the conduit for the energy exchange between them.

Let $P_{\xi}$ be the power or energy flow delivered by the noise force
\begin{equation}
	P_{\xi}(\tau)=\frac{e}{2}\,\langle\bigl\{\hat{\phi}_{h}(\mathbf{z},\tau),\,\dot{\hat{\chi}}(\tau)\bigr\}\rangle\,,
\end{equation}
and $P_{\gamma}$ be the power delivered by the backaction of radiation,
\begin{align}
	P_{\gamma}(\tau)&=\frac{e^{2}}{2}\int_{0}^{\tau}\!ds\;G_{R,\,0}^{(\phi)}(\mathbf{z},t;\mathbf{z},s)\langle\bigl\{\hat{\chi}(s),\,\dot{\hat{\chi}}(\tau)\bigr\}\rangle+\cdots=\frac{e^{2}}{2}\int_{0}^{t}\!ds\;G_{R,\,0}^{(\phi)}(\mathbf{z},t;\mathbf{z},s)\frac{d}{d\tau}G_{H}^{(\chi)}(s,\tau)+\cdots\,,
\end{align}
where $\cdots$ represents contributions associated with frequency renormalization, then the sum of both powers, the net energy exchange between the oscillator and the bath, is given by
\begin{equation}
	P_{\xi}(\tau)+P_{\gamma}(\tau)=\frac{e}{2}\,\langle\bigl\{\hat{\phi}(\mathbf{z},\tau),\,\dot{\hat{\chi}}(\tau)\bigr\}\rangle+\cdots\,.
\end{equation}
We now use the functional method to show that
\begin{equation}\label{E:fgbsrhers}
	\langle\bigl\{\hat{\phi}(\mathbf{z},\tau),\,\dot{\hat{\chi}}(\tau)\bigr\}\rangle=e\int_{0}^{\tau}\!ds\;\biggl\{\frac{d}{d\tau}G_{R}^{(\chi)}(\tau,s)\,G_{H,\,\beta}^{(\phi)}(s,\tau)+G_{R,\,0}^{(\phi)}(\tau-s)\frac{d}{d\tau}G_{H}^{(\chi)}(s,\tau)\biggr\}\,,
\end{equation}
such that 
\begin{align}
	P_{\xi}(\tau)&=e^{2}\int_{0}^{t}\!ds\;\frac{d}{d\tau}G_{R}^{(\chi)}(\tau,s)\,G_{H,\,\beta}^{(\phi)}(s,\tau)\,,\label{E:gkdjersaf}\\
	P_{\gamma}(\tau)&=\frac{e^{2}}{2}\int_{0}^{\tau}\!ds\;G_{R,\,0}^{(\phi)}(\mathbf{z},t;\mathbf{z},s)\frac{d}{d\tau}G_{H}^{(\chi)}(s,\tau)+\cdots\,.\label{E:gkdjersag}
\end{align}
This expression holds quite generally without resort to the perturbative expansion. Here we stress that $G^{(\phi)}$ denotes the two-point function of the free field and $G^{(\chi)}$ the two-point function for the full oscillator dynamics, including backactions from the field. In addition, under appropriate conditions, the net energy exchange will approach zero when the dynamics of the anharmonic oscillator is relaxed to an equilibrium state.

We can start from \eqref{E:gbkdue} with $\mathcal{Z}$ replaced by the generating functional $\mathcal{Z}_{V}$ of the anharmonic oscillator in \eqref{E:dvjgrtd}. In this case it is easier to use the functional derivatives with respect to $j_{\pm}$ rather than $j_{q}$, $j_{r}$. Thus we have, for $0<\tau,\tau'<t$,
\begin{align}
	\langle\hat{\phi}(\mathbf{z},\tau)\hat{\chi}(\tau')\rangle\,\mathcal{Z}_{V}&=e\int_{0}^{t}\!ds\;\biggl\{\frac{1}{2}\,G_{R,\,0}^{(\phi)}(s-\tau)\biggl[\frac{\delta^{2}}{i^{2}\delta j_{+}(\tau')\delta j_{+}(s)}+\frac{\delta^{2}}{i^{2}\delta j_{+}(\tau')\delta j_{-}(s)}\biggr]\biggr.\notag\\
	&\qquad\qquad\qquad+\frac{1}{2}\,G_{R,\,0}^{(\phi)}(\tau-s)\biggl[\frac{\delta^{2}}{i^{2}\delta j_{+}(\tau')\delta j_{+}(s)}-\frac{\delta^{2}}{i^{2}\delta j_{+}(\tau')\delta j_{-}(s)}\biggr]\notag\\
	&\qquad\qquad\qquad\qquad\qquad+\biggl.i\,G_{H,\,\beta}^{(\phi)}(\tau,s)\biggl[\frac{\delta^{2}}{i^{2}\delta j_{+}(\tau')\delta j_{+}(s)}+\frac{\delta^{2}}{i^{2}\delta j_{+}(\tau')\delta j_{-}(s)}\biggr]\biggr\}\,\mathcal{Z}_{V}\notag\\
	&=e\int_{0}^{t}\!ds\;\biggl\{-\frac{i}{2}\,G_{R,\,0}^{(\phi)}(s-\tau)\Bigl[G_{F}^{(\chi)}(\tau',s)-G_{<}^{(\chi)}(\tau',s)\Bigr]\biggr.\notag\\
	&\qquad\qquad\qquad-\frac{i}{2}\,G_{R,\,0}^{(\phi)}(\tau-s)\Bigl[G_{F}^{(\chi)}(\tau',s)+G_{<}^{(\chi)}(\tau',s)\Bigr]\notag\\
	&\qquad\qquad\qquad\qquad\qquad+\biggl.G_{H,\,\beta}^{(\phi)}(\tau,s)\Bigl[G_{F}^{(\chi)}(\tau',s)-G_{<}^{(\chi)}(\tau',s)\Bigr]\biggr\}\,\mathcal{Z}_{V}\,.\label{E:rjhsfds}
\end{align}
Now we will use the identities 
\begin{align}
	G_{F}(t,t')&=\frac{1}{2}\,G_{R}(t,t')+\frac{1}{2}\,G_{R}(t',t)+i\,G_{H}(t,t')\,,\\
	G_{F}(t,t')&+G_{<}(t,t')=G_{R}(t',t)+2i\,G_{H}(t,t')\,,
\end{align}
where, for the case of $\hat{\chi}$, we define the various two-point functions by
\begin{align*}
	G_{R}^{(\chi)}(t,t')&=i\,\theta(t-t')\,\langle\bigl[\chi(t),\chi(t')\bigr]\rangle\,,\qquad\qquad\qquad\qquad\qquad G_{H}^{(\chi)}(t,t')=\frac{1}{2}\,\langle\bigl\{\chi(t),\chi(t')\bigr\}\rangle\,,\\
	G_{F}^{(\chi)}(t,t')&=i\,\theta(t-t')\,\langle\chi(t)\chi(t')\rangle+i\,\theta(t'-t)\,\langle\chi(t')\chi(t)\rangle\,.
\end{align*}
Eq.~\eqref{E:rjhsfds} then becomes
\begin{align}
	\langle\hat{\phi}(\mathbf{z},\tau)\hat{\chi}(\tau')\rangle&=e\int_{0}^{t}\!ds\;\biggl\{G_{R}^{(\chi)}(\tau',s)\,G_{H,\,\beta}^{(\phi)}(s,\tau)+G_{R,\,0}^{(\phi)}(\tau-s)G_{H}^{(\chi)}(s,\tau')\biggr.\notag\\
	&\qquad\qquad\qquad\qquad-\biggl.\frac{i}{2}\,G_{R}^{(\chi)}(\tau',s)G_{R,\,0}^{(\phi)}(s-\tau)-\frac{i}{2}\,G_{R,\,0}^{(\phi)}(\tau-s)G_{R}^{(\chi)}(s,\tau')\biggr\}\,,
\end{align}
where $G_{R,\,0}^{(\phi)}$ represents the retarded Green's function of the free linear scalar field, while $G_{R}^{(\chi)}$denotes the full retarded Green's function of the nonlinear oscillator. We use $G_{R,\,0}^{(\chi)}$ for the corresponding retarded Green's function of the linear oscillator, or the zeroth-order contribution of the anharmonic oscillator.

Thus, in the coincident limit $\tau'\to\tau$, we obtain
\begin{align}
	\langle\hat{\phi}(\mathbf{z},\tau)\dot{\hat{\chi}}(\tau)\rangle&=e\int_{0}^{t}\!ds\;\biggl\{\frac{d}{d\tau}G_{R}^{(\chi)}(\tau,s)\,G_{H,\,\beta}^{(\phi)}(s,\tau)+G_{R,\,0}^{(\phi)}(\tau-s)\frac{d}{d\tau}G_{H}^{(\chi)}(s,\tau)\biggr.\notag\\
	&\qquad\qquad\qquad\qquad\qquad-\biggl.\frac{i}{2}\,\frac{d}{d\tau}G_{R}^{(\chi)}(\tau,s)G_{R,\,0}^{(\phi)}(s-\tau)-\frac{i}{2}\,G_{R,\,0}^{(\phi)}(\tau-s)\frac{d}{d\tau}G_{R}^{(\chi)}(s,\tau)\biggr\}\notag\\
	&=e\int_{0}^{t}\!ds\;\biggl\{\frac{d}{d\tau}G_{R}^{(\chi)}(\tau,s)\,G_{H,\,\beta}^{(\phi)}(s,\tau)+G_{R,\,0}^{(\phi)}(\tau-s)\frac{d}{d\tau}G_{H}^{(\chi)}(s,\tau)\biggr\}\,,\label{E:dbjrhtuwr}
\end{align}
because in the last two terms of the first line on the righthand side, according to the definition of the retarded Green's function, the variable $s$ is both greater and smaller than $\tau$, implying that $\tau=s$ and one of the retarded functions must be zero because $G_{R}(\tau,\tau)=0$ by definition. Therefore we have shown \eqref{E:fgbsrhers}, and obtained a general expression for the net energy exchange between the anharmonic oscillator and the bath field in terms of the two-point functions of the anharmonic oscillator and those of the free scalar field. It is interesting to know that in general the anharmonic oscillator is not a Gaussian system, so two-point functions or the second moments alone are not sufficient to describe its full statistics. Eq.~\eqref{E:gkdjersaf} or \eqref{E:dbjrhtuwr} will be used to find the rate of energy exchange between the anharmonic oscillator and the field bath. We will argue that under certain conditions, the net energy exchange will vanish when the motion of the anharmonic oscillator reaches equilibration. This will in turn imply a fluctuation-dissipation relation. The converse also holds.

In the special case of the linear oscillator, its retarded Green's function is stationary, and we can show that its Hadamard function in general is not stationary, but at late times the nonstationary component of the Hadamard function will be exponentially suppressed such that the Hadamard function becomes stationary. This allows us to show the existence of an FDR for the linear oscillator, which in turn implies that \eqref{E:fgbsrhers} vanishes for the linear oscillator coupled to the field bath at any coupling strength. It allows for the power balance and the existence of a stable steady state. However, we can not draw such general conclusions for the anharmonic oscillator. Since in general the driven, damped anharmonic oscillator does not necessarily have a unique stable steady state and quantum chaos can emerge, it is futile to pursue the same line of argument valid for the linear oscillator, and seek a general proof of the power balance or existence of the steady state for the anharmonic oscillator. We need stronger constraints on the configurations of the anharmonic oscillator for a steady state to exist. Thus we will use a weaker argument for \eqref{E:fgbsrhers}.

Before we proceed, we comment on the construction of the power operator by the canonical operator approach for the case of an anharmonic oscillator. For the linear oscillator we use the symmetric ordering to write down the operator for power, by which we then compute its expectation value. This arrangement becomes inappropriate in the context of the nonlinear oscillator. Take the quartic potential for the oscillator as an example. The power associated with the first two terms in \eqref{E:djsh}, according to the symmetric ordering, is given by
\begin{align}
	\frac{1}{2}\,\langle\bigl\{\ddot{\hat{\chi}}(t)+\omega^{2}\hat{\chi}(t),\,\dot{\hat{\chi}}(t)\bigr\}\rangle=\frac{d}{dt}\Bigl[\frac{m}{2}\,\dot{\hat{\chi}}^{2}(t)+\frac{m\omega^{2}}{2}\,\hat{\chi}^{2}(t)\Bigr]\,,
\end{align}
but for the power associated with the nonlinear restoring force, we note
\begin{equation}
	\frac{1}{2}\,\langle\bigl\{\hat{\chi}^{3}(t),\,\dot{\hat{\chi}}(t)\bigr\}\rangle\neq\frac{1}{4}\frac{d}{dt}\hat{\chi}^{4}(t)\,,
\end{equation}
whose righthand side in fact is
\begin{equation}
	\frac{1}{4}\frac{d}{dt}\hat{\chi}^{4}(t)=\frac{1}{4}\bigl[\hat{\chi}^{3}(t)\dot{\hat{\chi}}(t)+\hat{\chi}^{2}(t)\dot{\hat{\chi}}(t)\hat{\chi}(t)+\hat{\chi}(t)\dot{\hat{\chi}}(t)\hat{\chi}^{2}(t)+\dot{\hat{\chi}}(t)\hat{\chi}^{3}(t)\bigr]=\bigl[\hat{\chi}^{3}(t)\dot{\hat{\chi}}(t)\bigr]_{\textsc{w}}\,,
\end{equation}
that is, the Weyl ordering or fully symmetrized ordering. This is the same as the symmetric ordering when there are only two operators involved. Thus, if we use the Weyl ordering to construct the power operator, we can rewrite the lefthand side of \eqref{E:djsh} as
\begin{equation}
	\frac{d}{dt}\Bigl\{\frac{m}{2}\,\dot{\hat{\chi}}^{2}(t)+\frac{m\omega_{\textsc{r}}^{2}}{2}\,\hat{\chi}^{2}(t)+\lambda\,V[\hat{\chi}(t)]\Bigr\}=P_{\xi}(\tau)+P_{\gamma}(\tau)\,,
\end{equation}
where we have introduced the renormalized frequency $\omega_{\textsc{r}}$.

Since it is not possible to prove the existence of a stable steady state for the general configurations of the anharmonic oscillator, {we take a step down, i.e., we consider only those cases when the dynamics of the anharmonic oscillator can indeed reach a stable steady state at late times.} Namely,the energy exchange between the oscillator and the field bath will come into equilibrium. That is, the rate of the energy exchange will become zero at late times, implying
\begin{align}
	\lim_{\tau\to\infty}P_{\xi}(\tau)+P_{\gamma}(\tau)&=\lim_{\tau\to\infty}e^{2}\int_{0}^{\tau}\!ds\;\biggl\{\frac{d}{d\tau}G_{R}^{(\chi)}(\tau,s)\,G_{H,\,\beta}^{(\phi)}(s,\tau)+G_{R,\,0}^{(\phi)}(\tau-s)\frac{d}{d\tau}G_{H}^{(\chi)}(s,\tau)\biggr\}+\cdots\notag\\
	&=0
\end{align}
after we have subtracted the contribution corresponding to frequency renormalization. We set forth to show that this energy balance implies a fluctuation-dissipation relation for the anharmonic oscillator.

We first deal with the frequency renormalization more explicitly by introducing a new kernel function $\Gamma_{R,\,0}^{(\phi)}(t)$ for the free field:
\begin{equation}\label{E:ithrhkfd}
	G_{R,\,0}^{(\phi)}(\mathbf{z},t;\mathbf{z},s)\equiv G_{R,\,0}^{(\phi)}(t-s)=\frac{d}{ds}\Gamma_{R,\,0}^{(\phi)}(t-s)\,.
\end{equation}
After an integration by parts, we find
\begin{align}
	\int_{0}^{\tau}\!ds\;G_{R,\,0}^{(\phi)}(\tau-s)\frac{d}{d\tau}G_{H}^{(\chi)}(s,\tau)&=\int_{0}^{\tau}\!ds\;\frac{d}{ds}\Gamma_{R,\,0}^{(\phi)}(\tau-s)\frac{d}{d\tau}G_{H}^{(\chi)}(s,\tau)\notag\\
	&=\Gamma_{R,\,0}^{(\phi)}(0)\frac{d}{d\tau}\biggl(\frac{\hat{\chi}^{2}(\tau)}{2}\biggr)-\Gamma_{R,\,0}^{(\phi)}(\tau)\frac{d}{d\tau}G_{H}^{(\chi)}(0,\tau)\\
	&\qquad\qquad\qquad\qquad\qquad\qquad\qquad-\int_{0}^{\tau}\!ds\;\Gamma_{R,\,0}^{(\phi)}(\tau-s)\frac{d^{2}}{ds\,d\tau}G_{H}^{(\chi)}(s,\tau)\,.\notag
\end{align}
For a scalar field bath, the kernel $\Gamma^{(\phi)}_{R,\,0}(t)$ is proportional to the delta function $\delta(t)$, so the second term on the righthand side vanishes. Now it is clear to see that the first term  corresponds to  frequency renormalization, and thus
\begin{equation}\label{E:gbrsijdfs}
	P_{\xi}(\tau)+P_{\gamma}(\tau)=e^{2}\int_{0}^{\tau}\!ds\;\biggl\{\frac{d}{d\tau}G_{R}^{(\chi)}(\tau,s)\,G_{H,\,\beta}^{(\phi)}(s,\tau)-\Gamma_{R,\,0}^{(\phi)}(\tau-s)\frac{d^{2}}{ds\,d\tau}G_{H}^{(\chi)}(s,\tau)\biggr\}\,.
\end{equation}
The assumed existence of the steady state implies that at late times the two-point functions of the oscillator will become  stationary since in the Heisenberg picture we have
\begin{align}
	\langle\hat{\chi}(t)\hat{\chi}(t')\rangle=\operatorname{Tr}_{\chi}\bigl\{\hat{\rho}(0)\hat{\chi}(t)\hat{\chi}(t')\bigr\}&=\operatorname{Tr}_{\chi}\bigl\{\hat{\rho}(t')\hat{U}^{-1}(t-t')\hat{\chi}(0)\hat{U}(t-t')\hat{\chi}(0)\bigr\}\,,\label{E:rksdheuris}
\end{align}
where $\rho(0)$ is the initial state of $\chi$ and $\hat{U}$ is the time evolution operator. If at late times the oscillator approaches a steady state, we expect that for sufficiently large $t$, $t'$, we have $\hat{\rho}(t')\simeq\hat{\rho}(t)=\text{const}$. This seems to suggest that \eqref{E:rksdheuris} becomes a function of $t-t'$. However, this property still can not allow us to write \eqref{E:gbrsijdfs} as
\begin{equation}\label{E:tbrtbee}
	P_{\xi}(\tau)+P_{\gamma}(\tau)\stackrel{?}{=}\,e^{2}\int_{0}^{\tau}\!ds\;\biggl\{\frac{d}{d\tau}G_{R}^{(\chi)}(\tau-s)\,G_{H,\,\beta}^{(\phi)}(\tau-s)-\Gamma_{R,\,0}^{(\phi)}(\tau-s)\frac{d^{2}}{ds\,d\tau}G_{H}^{(\chi)}(\tau-s)\biggr\}\,,
\end{equation}
for sufficiently large $\tau$. Although the nonstationary component of the oscillator's two-point functions decay to zero when both time arguments are large, its contribution to the integral in \eqref{E:gbrsijdfs} may remain substantial if it does not decay fast enough. In the case of the linear oscillator, the contribution of the nonstationary component diminishes exponentially fast if the nonstationarity results from the bilinear interaction with the field bath, not from the initial nonstationary state like a squeezed state. We can show that the equal sign in \eqref{E:tbrtbee} holds for a linear oscillator.

For the nonlinear oscillator, we need a stronger assumption:: nsider the case that the nonstationary component of the oscillator's two-point function does not contribute to the integral in \eqref{E:gbrsijdfs}, thus validating \eqref{E:tbrtbee}. Under this assumption, we can write the integral in \eqref{E:tbrtbee} as
\begin{align}
	&\quad\int_{0}^{\tau}\!ds\;\biggl\{\frac{d}{d\tau}G_{R}^{(\chi)}(\tau-s)\,G_{H,\,\beta}^{(\phi)}(\tau-s)+\Gamma_{R,\,0}^{(\phi)}(\tau-s)\frac{d^{2}}{d\tau^{2}}G_{H}^{(\chi)}(\tau-s)\biggr\}\notag\\
	&=\int_{-\infty}^{\tau}\!dy\;\biggl\{\frac{d}{dy}G_{R}^{(\chi)}(y)\,G_{H,\,\beta}^{(\phi)}(y)+\Gamma_{R,\,0}^{(\phi)}(y)\frac{d^{2}}{dy^{2}}G_{H}^{(\chi)}(y)\biggr\}\,,&y&=\tau-s\,,
\end{align}
due to the retardation property of the kernel function. In the large $\tau$ limit, we have
\begin{align}
	&\quad\lim_{\tau\to\infty}\int_{-\infty}^{\tau}\!dy\;\biggl\{\frac{d}{dy}G_{R}^{(\chi)}(y)\,G_{H,\,\beta}^{(\phi)}(y)+\Gamma_{R,\,0}^{(\phi)}(y)\frac{d^{2}}{dy^{2}}G_{H}^{(\chi)}(y)\biggr\}\notag\\
	&=i\int_{-\infty}^{\infty}\!\frac{d\kappa}{2\pi}\;\kappa\,\biggl\{\overline{G}_{R}^{(\chi)*}\!(\kappa)\,\overline{G}_{H,\,\beta}^{(\phi)}(\kappa)-\overline{G}_{R,\,0}^{(\phi)*}(\kappa)\,\overline{G}_{H}^{(\chi)}(\kappa)\biggr\}\,,\label{E:gbkfjbrj}
\end{align}
since $\overline{G}_{R,\,0}^{(\phi)}(\kappa)=i\,\kappa\,\overline{\Gamma}_{R,\,0}^{(\phi)}(\kappa)$. The kernel functions of the free field satisfy the FDR 
\begin{equation}\label{E:rhfgjdh}
	\overline{G}_{H,\,\beta}^{(\phi)}(\kappa)=\coth\frac{\beta\kappa}{2}\,\operatorname{Im}\overline{G}_{R,\,0}^{(\phi)}(\kappa)\,,
\end{equation}
if the field is initially in the thermal state. This implies that \eqref{E:gbkfjbrj} becomes
\begin{align}
	&=\int_{-\infty}^{\infty}\!\frac{d\kappa}{2\pi}\;\kappa\,\biggl\{\coth\frac{\beta\kappa}{2}\,\operatorname{Im}\overline{G}_{R}^{(\chi)}\!(\kappa)-\overline{G}_{H}^{(\chi)}(\kappa)\biggr\}\,\operatorname{Im}\overline{G}_{R,\,0}^{(\phi)}(\kappa)\,,
\end{align}
where we have used the properties that $\overline{G}_{H}(\kappa)$, $\operatorname{Re}\overline{G}_{R}(\kappa)$ are even with respect to $\kappa$ but $\operatorname{Im}\overline{G}_{R}(\kappa)$ is an odd function of $\kappa$. Thus the condition that at late times $P_{\xi}+P_{\gamma}$ vanishes implies an FDR for the anharmonic oscillator
\begin{equation}\label{E:djgeresfjk}
	\overline{G}_{H}^{(\chi)}(\kappa)=\coth\frac{\beta\kappa}{2}\,\operatorname{Im}\overline{G}_{R}^{(\chi)}\!(\kappa)\,.
\end{equation}
This is a nonperturbative result but it requires two rather strong assumptions: 1) a stable steady state exists at late times, and 2) the nonstationary component of the anharmonic oscillator's two-point function has negligible contribution to the integral in \eqref{E:gbrsijdfs} at late times. These two assumptions can be directly shown to be true for the linear oscillator,   but it is not clear yet under what conditions they also hold for the anharmonic oscillator. {Nonetheless, in the conceptual framework of quantum open systems  the FDR \eqref{E:djgeresfjk} derived in the context of nonequilibrium dynamics (as opposed to linear response) for an anharmonic oscillator coupled to a field bath registers a deep connection  between equilibration of the system with its environment,  balance of energy flow and stationarity in the late-time dynamics. In the next section, we shall use perturbative calculations to ascertain }  these two assumptions.

If the assumption that the nonstationary component of the two-point function of the nonlinear oscillator vanishes sufficiently fast is lifted, then we do not have \eqref{E:tbrtbee}. Thus the existence of the steady state at late times will instead imply an integral FDR of the anharmonic oscillator like
\begin{equation}
	\lim_{\tau\to\infty}\int_{0}^{\tau}\!ds\;\biggl\{\frac{d}{d\tau}G_{R}^{(\chi)}(\tau,s)\,G_{H,\,\beta}^{(\phi)}(s,\tau)-\Gamma_{R,\,0}^{(\phi)}(\tau-s)\frac{d^{2}}{ds\,d\tau}G_{H}^{(\chi)}(s,\tau)\biggr\}=0\,,
\end{equation}
in the time domain, togeher with \eqref{E:rhfgjdh}, rather than an algebraic relation \eqref{E:djgeresfjk} in the frequency domain.

\section{Energy Flow Balance between an Anharmonic Oscillator and its Quantum Field Bath - Perturbative Arguments}\label{S:ierbdsfwea}

We now illustrate the perturbative calculations established above by examining the condition for the equilibration of a weakly anharmonic oscillator interacting with a quantum scalar field.  An important signature for the presence of such an equilibrium state in the motion of the anharmonic oscillator  coupled with a quantum field is that the rate of energy exchange between the oscillator and the field must be balanced such that the net energy flow approaches zero at late times. This condition is not easy to verify for an anharmonic oscillator due to the lack of complete late-time analytical expressions of its observables. In general, results based on the perturbative expansion are not reliable at large evolution time, especially when the nonliner system is driven by a periodic source. The error accumulation, the secular evolution, and the onset of chaos phenomena associated with the nonlinear system often limit the perturbative calculations to the short time regime. Nonetheless,  perturbative treatments may still be applicable to some configurations, restrictive as they may be. One such configuration is the small-amplitude oscillation of a weakly anharmonic oscillator coupled to a low temperature quantum-field bath, in which the small-amplitude and the weak nonlinearity warrants a perturbative treatment, and the backaction from the field bath  induces only a weak stochastic noise and damping. If the anharmonic potential is such that $\chi=0$ remains the unique global minimum of the potential and the other local minima are located far away from $\chi=0$ whereby tunneling is suppressed,  the damping dynamics in the system resulting from its interaction with the field would be enough to confine the late-time motion of such a configuration around the global minimum at $\chi=0$.

This may not be the most interesting scenario, but it shows that even perturbative treatment may provide some meaningful description of the late-time dynamics. With the help of the functional method discussed in Appendix~\ref{E:gjerhs}, we find  the first-order correction to $\langle\hat{\phi}(\mathbf{z},\tau)\dot{\hat{\chi}}(\tau)\rangle$ given by
\begin{align}
	\langle\hat{\phi}(\mathbf{z},\tau)\dot{\hat{\chi}}(\tau)\rangle^{(1)}=e\int_{0}^{t}\!ds\;\biggl\{G_{R,\,0}^{(\phi)}(\tau-s)\,\Bigl[\frac{d}{d\tau}G_{H,1}^{(\chi)}(s,\tau)\Bigr]+\Bigl[\frac{d}{d\tau}G_{R,1}^{(\chi)}(\tau,s)\Bigr]G_{H,0}^{(\phi)}(s,\tau)\biggr\}\,,\label{E:fgjkhraa}
\end{align}
so that from \eqref{E:fgbsrhers} we have
\begin{align}\label{E:erkhbdsd}
	P_{\xi}^{(1)}(\tau)+P_{\gamma}^{(1)}(\tau)&=e^{2}\int_{0}^{\tau}\!ds\;\biggl\{G_{R,\,0}^{(\phi)}(\tau-s)\,\Bigl[\frac{d}{d\tau}G_{H,1}^{(\chi)}(s,\tau)\Bigr]+\Bigl[\frac{d}{d\tau}G_{R,1}^{(\chi)}(\tau,s)\Bigr]G_{H,0}^{(\phi)}(s-\tau)\biggr\}+\cdots\,.
\end{align}
with
\begin{align}
	P_{\xi}^{(1)}(\tau)&=e^{2}\int_{0}^{\tau}\!ds\;\Bigl[\frac{d}{d\tau}G_{R,1}^{(\chi)}(\tau,s)\Bigr]G_{H,0}^{(\phi)}(s-\tau)\,,\label{E:ruthrdb1}\\
	P_{\gamma}^{(1)}(\tau)&=e^{2}\int_{0}^{\tau}\!ds\;G_{R,\,0}^{(\phi)}(\tau-s)\,\Bigl[\frac{d}{d\tau}G_{H,1}^{(\chi)}(s,\tau)\Bigr]+\cdots\,,\label{E:ruthrdb2}
\end{align}
where $\cdots$ represents contributions related to the frequency renormalization, and the superscript $(1)$ denotes the correction of first order in the self-coupling constant $\lambda$ in the anharmonic potential. Note that we have implicitly assumed that the initial state of the massless scalar field $\phi$ is a stationary state.

Although  we have shown that the first-order correction of the kernel functions $G_{R,1}^{(\chi)}(\tau,\tau')$ and $G_{H,1}^{(\chi)}(\tau,\tau')$ of the nonlinear oscillator become stationary when $\tau$ and $\tau'$ are sufficiently large, we still cannot replace $G_{R,1}^{(\chi)}(\tau,s)$ and $G_{H,1}^{(\chi)}(s,\tau)$ in the integrals by $G_{R,1}^{(\chi)}(\tau-s)$ and $G_{H,1}^{(\chi)}(s-\tau)$ in the late-time limit $\tau\to\infty$ because $s$ ranges from 0 to $\tau\to\infty$. We need to show that the contributions from their nonstationary components to the integral are negligible. For $P_{\gamma}^{(1)}$, the arguments is rather straightforward. Since in~\cite{NENL1} we have argued that for sufficiently large $\tau$, the nonstationary component of $G_{H,1}^{(\chi)}(s,\tau)$ will contain a factor like $e^{-\gamma s}$ and since we observe that for a massless scalar field $\phi$, its retarded function $G_{R,0}^{(\phi)}(\tau-s)$ drops to zero rapidly when $s$ deviates from $\tau$, we conclude that the dominant contribution of the integral \eqref{E:ruthrdb2} will come from the values of $s$ in the vicinity of $\tau$. Thus in the limit $\tau\to\infty$, we can write $G_{H,1}^{(\chi)}(s,\tau)$ in \eqref{E:ruthrdb2} as approaching $G_{H,1}^{(\chi)}(s-\tau)$.

For $P_{\xi}^{(1)}$, let us examine the contribution from the nonstationary component of $G_{R,1}^{(\chi)}(\tau,\tau')$. We first observe that by construction in \eqref{E:oerwpijen2}, $G_{R,1}^{(\chi)}(\tau,\tau,)$ is given by
\begin{align}
	G_{R,1}^{(\chi)}(\tau,\tau')&=-\frac{\lambda}{2!}\int_{0}^{\tau}\!ds\;G_{R,0}^{(\chi)}(\tau-s)G_{H,0}^{(\chi)}(s,s)G_{R,0}^{(\chi)}(s-\tau')\\
	&=-\frac{\lambda}{4\Omega^{2}}\int_{0}^{\tau}\!ds\;e^{-\gamma(\tau-\tau')}\Bigl[\cos\Omega(\tau+\tau'-2s)-\cos\Omega(\tau-\tau')\Bigr]G_{H,0}^{(\chi)}(s,s)\,,\notag
\end{align}
where we have substituted the expressions of $G_{R,0}^{(\chi)}(\tau-s)$. It is useful to write it as the sum of two integrals $I_{1}$ and $I_{2}$
\begin{align*}
	I_{1}&=\frac{\lambda}{4\Omega^{2}}\,e^{-\gamma(\tau-\tau')}\cos\Omega(\tau-\tau')\int_{\tau'}^{\tau}\!ds\;G_{H,0}^{(\chi)}(s,s)\,,\\
	I_{2}&=-\frac{\lambda}{4\Omega^{2}}\,e^{-\gamma(\tau-\tau')}\int_{\tau'}^{\tau}\!ds\;\cos\Omega(\tau+\tau'-2s)\,G_{H,0}^{(\chi)}(s,s)\,.
\end{align*}
We note~\cite{NENL1} that in general $G_{H,0}^{(\chi)}(s,s)$ take the form
\begin{equation}
	G_{H,0}^{(\chi)}(s,s)=\text{const.}+e^{-\gamma s}\bigl(\cdots\bigr)+e^{-2\gamma s}\bigl(\cdots\bigr)\,,
\end{equation}
where terms in $(\cdots)$ are sinusoidal in $s$, i.e., $e^{\pm i\omega s}$ with some constant $\omega\in\mathbb{R}$, and ``const." represents terms independent of $s$. Since we observe that generically
\begin{align*}
	&\int_{\tau'}^{\tau}\!ds=\tau-\tau'\,,\qquad\qquad\qquad\qquad\qquad\qquad\qquad\int_{\tau'}^{\tau}\!ds\;e^{-\gamma s}e^{i\omega s}=-\frac{e^{-\gamma\tau+i\,\omega \tau}}{\gamma-i\,\omega}+\frac{e^{-\gamma\tau'+i\,\omega \tau'}}{\gamma-i\,\omega}\,,\\
	&\int_{\tau'}^{\tau}\!ds\;\cos\Omega(\tau+\tau'-2s)=\frac{1}{\Omega}\,\sin\Omega(\tau-\tau')\,,\\
	&\int_{\tau'}^{\tau}\!ds\;e^{-\gamma s}e^{i\omega s}\cos\Omega(\tau+\tau'-2s)=A\,e^{\gamma\tau+i\,\omega\tau}g(\tau-\tau')+B\,e^{\gamma\tau'+i\,\omega\tau'}g(\tau-\tau')
\end{align*}
where $A$, $B$ are $\tau$, $\tau'$-independent constants, and $g(s)$ is some bounded sinusoidal function of $s$. The point is that the nonstationary terms in $G_{R,1}^{(\chi)}$ are always exponentially smaller than the stationary terms. Thus the contribution from the nonstationary terms of $G_{R,1}^{(\chi)}$ to the integral in the definition $P_{\xi}^{(1)}$ for $\tau\to\infty$ will be typically likewise smaller than the contribution from the stationary terms of $G_{R,1}^{(\chi)}$. This implies that we can drop the nonstationary terms in $G_{R,1}^{(\chi)}(\tau,\tau')$ with negligible errors. In other word, we now can write $G_{R,1}^{(\chi)}(\tau,\tau')=G_{R,1}^{(\chi)}(\tau-\tau')$ in \eqref{E:ruthrdb1} such that
\begin{equation}\label{E:dgjverhs}
	P_{\xi}^{(1)}(\tau)=e^{2}\int_{0}^{\tau}\!ds\;\Bigl[\frac{d}{d\tau}G_{R,1}^{(\chi)}(\tau-s)\Bigr]G_{H,0}^{(\phi)}(s-\tau)\,,
\end{equation}
for large $\tau$, and in this limit the net power becomes
\begin{align}\label{E:erkhebdsd}
	P_{\xi}^{(1)}(\tau)+P_{\gamma}^{(1)}(\tau)=e^{2}\int_{0}^{\tau}\!ds\;\biggl\{G_{R,\,0}^{(\phi)}(\tau-s)\,\Bigl[\frac{d}{d\tau}G_{H,1}^{(\chi)}(s-\tau)\Bigr]+\Bigl[\frac{d}{d\tau}G_{R,1}^{(\chi)}(\tau-s)\Bigr]G_{H,0}^{(\phi)}(s-\tau)\biggr\}+\cdots\,.
\end{align}
Following the arguments between \eqref{E:ithrhkfd} and \eqref{E:djgeresfjk} and the perturbative FDR \eqref{E:rturtdfjw} for the nonlinear oscillator, we conclude that the net power will vanish at late times $\tau\to\infty$. The reasoning used here offer some glimpse into the plausibility of the assumption used in the nonperturbative arguments, regarding the negligible contribution from the nonstationary component of the kernel functions of the nonlinear oscillator to the energy exchange with the environment scalar field.\\

\noindent {\bf Acknowledgments}  This work makes use of results obtained when both authors visited the Center for Particle Physics and Field Theory of Fudan University in 2013 and 2014. It is further developed when JTH visited the Maryland Center for Fundamental Physics at the University of Maryland and when BLH visited the National Center for Theoretical Sciences in Hsinchu, Taiwan in 2019.    



\appendix

\section{Derivation of Energy Flow $P_{\xi}$}\label{S:powndkfgwsd}\label{E:gjerhs}

To illustrate the functional perturbative method we calculate the power $P_{\xi}$ delivered by a massless quantum scalar field $\hat{\phi}(\mathbf{x},t)$  to the  {linear (harmonic)} oscillator $\hat{\chi}(t)$, which can be viewed as the internal degrees of freedom of  an Unruh-DeWitt detector.

The power delivered by the free field at time $\tau$ is defined by
\begin{equation}\label{E:rbhgd}
	P_{\xi}(\tau)=e\langle\hat{\phi}_{h}(\mathbf{z},\tau)\hat{\chi}(\tau)\rangle
\end{equation}
where the external degree of freedom $\mathbf{z}$ of the detector gives the fixed location of the detector, and $\hat{\phi}_{h}(\mathbf{z},t)$ is the free field component, that is, the homogeneous solution of the field equation.  {Although \eqref{E:rbhgd} is the real part of the coincident limit,}
\begin{equation}
	e\,\lim_{\tau'\to\tau}\frac{d}{d\tau'}\langle\hat{\phi}_{h}(\mathbf{z},\tau)\hat{\chi}(\tau')\rangle\,.
\end{equation}
we will use the functional method to derive the expressions for
\begin{equation}\label{E:bghseht}
	\langle\hat{\phi}(\mathbf{z},\tau)\hat{\chi}(\tau')\rangle\,.
\end{equation}
{in terms of two-point functions of the oscillator and the free field}. Note that  {\eqref{E:bghseht} involves the full interacting field $\hat{\phi}(\mathbf{z},t)$,} instead of the free field $\hat{\phi}_{h}(\mathbf{z},t)$, so the result  {will contain an additional contribution from the radiation field emitted by} the evolving oscillator. For a linear oscillator it is pretty easy to distinguish the contribution of the free field from that of the radiation field\footnote{{From Eq.~\eqref{E:vher3}, we observe that the first term on its right-hand side, the homogeneous solution of the field equation~\eqref{E:vher2}, describes the free quantum field, while the second term, the inhomogeneous solution, is a radiation field, i.e. Li\'enard Wiechert potential~\cite{JD75} emitted by the oscillator at $\mathbf{z}$. More discussions can be found in~\cite{QRad19}.}}.

Since the expectation value \eqref{E:bghseht} contains the field operator, it is not obvious a priori how to apply the functional method to the generating functional we have in Sec.~\ref{S:drnkw} to compute the expectation value. We return to the starting point, noting that \eqref{E:bghseht} means
\begin{equation}
	\langle\hat{\phi}(\mathbf{z},\tau)\hat{\chi}(\tau')\rangle=\operatorname{Tr}_{\chi\phi}\Bigl\{\hat{\rho}_{\chi\phi}(t)\hat{\phi}(\mathbf{z},\tau)\hat{\chi}(\tau')\Bigr\}
\end{equation}
where $\hat{\rho}_{\chi\phi}$ is the density operator of the whole system, $0<\tau,\,\tau'<t$ and $t$ can be taken to $+\infty$ for convenience. Introducing a path integral representation, we have
\begin{align}
	&\quad\langle\hat{\phi}(\mathbf{z},\tau)\hat{\chi}(\tau')\rangle_{j}\,\mathcal{Z}[j;t)\notag\\
	&=\int_{-\infty}^{\infty}\!d\chi_{b}d\chi'_{b}\;\delta(\chi_{b}-\chi'_{b})\int_{-\infty}^{\infty}\!d\chi_{a}d\chi'_{a}\;\rho_{\chi}(\chi_{a},\chi'_{a},0)\notag\\
	&\qquad\qquad\qquad\times\int_{\chi_{a}}^{\chi_{b}}\!\mathcal{D}\chi_{+}\!\int_{\chi'_{a}}^{\chi'_{b}}\!\mathcal{D}\chi_{-}\;\chi_{+}(\tau')\,\exp\Bigl(i\,S_{\chi}[\chi_{+},j_{+}]-i\,S_{\chi}[\chi_{-},j_{-}]\Bigr)\notag\\
	 &\qquad\qquad\qquad\times\int_{-\infty}^{\infty}\!d\phi_{b}d\phi'_{b}\;\delta(\phi_{b}-\phi'_{b})\int_{-\infty}^{\infty}\!d\phi_{a}d\phi'_{a}\;\rho_{\phi}(\phi_{a},\phi'_{a},0)\notag\\
	 &\qquad\qquad\qquad\times\int_{\phi_{a}}^{\phi_{b}}\!\mathcal{D}\phi_{+}\!\int_{\phi'_{a}}^{\phi'_{b}}\!\mathcal{D}\phi_{-}\;\phi_{+}(\mathbf{z},\tau)\,\exp\Bigl(i\,S_{I}[\chi_{+},\phi_{+}]-i\,S_{I}[\chi_{-},\phi_{-}]+i\,S_{\phi}[\phi_{+}]-i\,S_{\phi}[\phi_{-}]\Bigr)\label{E:gbretuer}\\
	 &=\frac{\delta}{i\,\delta j_{+}(\tau')}\int_{-\infty}^{\infty}\!d\chi_{b}d\chi'_{b}\;\delta(\chi_{b}-\chi'_{b})\int_{-\infty}^{\infty}\!d\chi_{a}d\chi'_{a}\;\rho_{\chi}(\chi_{a},\chi'_{a},0)\notag\\
	 &\qquad\qquad\qquad\times\int_{\chi_{a}}^{\chi_{b}}\!\mathcal{D}\chi_{+}\!\int_{\chi'_{a}}^{\chi'_{b}}\!\mathcal{D}\chi_{-}\;\exp\Bigl(i\,S_{\chi}[\chi_{+},j_{+}]-i\,S_{\chi}[\chi_{-},j_{-}]\Bigr)\notag\\
	 &\qquad\qquad\qquad\times\frac{1}{e}\frac{\delta}{i\,\delta\chi_{+}(\tau)}\int_{-\infty}^{\infty}\!d\phi_{b}d\phi'_{b}\;\delta(\phi_{b}-\phi'_{b})\int_{-\infty}^{\infty}\!d\phi_{a}d\phi'_{a}\;\rho_{\phi}(\phi_{a},\phi'_{a},0)\notag\\
	 &\qquad\qquad\qquad\times\int_{\phi_{a}}^{\phi_{b}}\!\mathcal{D}\phi_{+}\!\int_{\phi'_{a}}^{\phi'_{b}}\!\mathcal{D}\phi_{-}\;\exp\Bigl(i\,S_{I}[\chi_{+},\phi_{+}]-i\,S_{I}[\chi_{-},\phi_{-}]+i\,S_{\phi}[\phi_{+}]-i\,S_{\phi}[\phi_{-}]\Bigr)\notag\\
	 &=\frac{\delta}{i\,\delta j_{+}(\tau')}\int_{-\infty}^{\infty}\!d\chi_{b}d\chi'_{b}\;\delta(\chi_{b}-\chi'_{b})\int_{-\infty}^{\infty}\!d\chi_{a}d\chi'_{a}\;\rho_{\chi}(\chi_{a},\chi'_{a},0)\notag\\
	 &\qquad\qquad\qquad\times\int_{\chi_{a}}^{\chi_{b}}\!\mathcal{D}\chi_{+}\!\int_{\chi'_{a}}^{\chi'_{b}}\!\mathcal{D}\chi_{-}\;\exp\Bigl(i\,S_{\chi}[\chi_{+},j_{+}]-i\,S_{\chi}[\chi_{-},j_{-}]\Bigr)\notag\\
	 &\qquad\qquad\qquad\times\biggl\{e\int_{0}^{t}\!ds\;\biggl[\frac{1}{2}\,G_{R,0}^{(\phi)}(s-\tau)\,q(s)+G_{R,0}^{(\phi)}(\tau-s)\,r(s)+i\,G_{H,\,\beta}^{(\phi)}(\tau,s)\,q(s)\biggr]\biggr\}\notag\\
	 &\qquad\qquad\qquad\times\exp\biggl\{i\,e^{2}\int_{0}^{t}\!ds\,ds'\;\Bigl[q(s)\,G_{R,0}^{(\phi)}(s,s')\,r(s')+\frac{i}{2}\,q(s)\,G_{H,\,\beta}^{(\phi)}(s,s')\,q(s')\Bigr]\biggr\}\notag\\
	&=-e\int_{0}^{t}\!ds\;\biggl\{\biggl[\frac{1}{4}\,G_{R,0}^{(\phi)}(s-\tau)\,\frac{\delta^{2}}{\delta j_{r}(\tau')\delta j_{r}(s)}+\frac{i}{2}\,G_{H,\,\beta}^{(\phi)}(\tau,s)\,\frac{\delta^{2}}{\delta j_{r}(\tau')\delta j_{r}(s)}\biggr]\biggr.\notag\\
	&\qquad\qquad\qquad\quad+\biggl[\frac{1}{2}\,G_{R,0}^{(\phi)}(\tau-s)\,\frac{\delta^{2}}{\delta j_{r}(\tau')\delta j_{q}(s)}+\frac{1}{2}\,G_{R,0}^{(\phi)}(s-\tau)\,\frac{\delta^{2}}{\delta j_{q}(\tau')\delta j_{r}(s)}\biggr.\notag\\
	&\qquad\qquad\qquad\qquad\qquad+\biggl.\biggl.i\,G_{H,\,\beta}^{(\phi)}(\tau,s)\,\frac{\delta^{2}}{\delta j_{q}(\tau')\delta j_{r}(s)}\biggr]+G_{R,0}^{(\phi)}(\tau-s)\,\frac{\delta^{2}}{\delta j_{q}(\tau')\delta j_{q}(s)}\biggr\}\times\mathcal{Z}[j_{r},j_{q};t)\,,\label{E:gbkdue}
\end{align}
where the oscillator action $S_{\chi}[\chi,j]$  contains the contributions from the external sources,
\begin{equation}
	S_{\chi}[\chi,j]=\int_{0}^{t}\!ds\;\Bigl\{\frac{m}{2}\Bigl[\dot{\chi}^{2}(s)-\omega^{2}\chi^{2}(s)\Bigr]-V[\chi(s)]+j(s)\chi(s)\Bigr\}\,,
\end{equation}
and $V[\chi]$ accounts for the nonlinear potential.  {Here, be reminded that $G_{R,0}^{(\phi)}$ and $G_{H,\,\beta}^{(\phi)}$ individually denotes the retarded Green's function and the Hadamard function of the \textit{free} field, while $G_{R}^{(\chi)}$ and $G_{H}^{(\chi)}$ are respectively the retarded Green's function and the Hadamard function of the  oscillator   interacting with the scalar field.}

Here Eq.~\eqref{E:gbkdue} indicates that the insertion of $\phi(\mathbf{z},\tau)$ is equivalent to taking an additional functional derivative of the combination
\begin{equation*}
	e\int_{0}^{t}\!ds\;\biggl[\frac{1}{2}\,G_{R,0}^{(\phi)}(s-\tau)\,\frac{\delta}{i\,\delta j_{r}(s)}+G_{R,0}^{(\phi)}(\tau-s)\,\frac{\delta}{i\,\delta j_{q}(s)}+i\,G_{H,\,\beta}^{(\phi)}(\tau,s)\,\frac{\delta}{i\,\delta j_{r}(s)}\biggr]
\end{equation*}
of the generating functional $\mathcal{Z}[j;t)$. Since it originates from the influence action,  we can infer that the terms involving the retarded Green's function of the free field $G_{R,0}^{(\phi)}$ are related to the radiation field, caused by the nontrivial motion of the detector, while that associated with the Hadamard function of the free field $G_{H,\,\beta}^{(\phi)}$  pertains to the {free field fluctuations}. The latter will be what we search for in calculating \eqref{E:bghseht} and the power delivered by the free field $\hat{\phi}_{h}(\mathbf{z},t)$.

Let us first consider the simpler case when the nonlinear potential $V[\chi]$ is absent. To evaluate \eqref{E:gbkdue}, we need the following identities from previous calculations
\begin{align}
	\frac{\delta^{2}\mathcal{Z}[j;t)}{\delta j_{q}(\tau)\,\delta j_{q}(\tau')}&=i\,\frac{\delta\,\Xi[j;\tau')}{\delta j_{q}(\tau)}\,\mathcal{Z}[j;t)-\Xi[j;\tau)\Xi[j;\tau')\mathcal{Z}[j;t)\,,\\
	\frac{\delta^{2}\mathcal{Z}[j;t)}{\delta j_{q}(\tau')\delta j_{r}(\tau)}&=i\,\frac{\delta\,\mathfrak{J}_{q}(\tau)}{\delta j_{q}(\tau')}\,\mathcal{Z}[j;t)-\mathfrak{J}_{q}(\tau)\,\Xi[j;\tau')\mathcal{Z}[j;t)\,,\\
	\frac{\delta^{2}\mathcal{Z}[j;t)}{\delta j_{q}(\tau)\,\delta j_{r}(\tau')}&=i\,\frac{\delta\,\mathfrak{J}_{q}(\tau')}{\delta j_{q}(\tau)}\,\,\mathcal{Z}[j;t)-\mathfrak{J}_{q}(\tau')\,\Xi[j;\tau)\mathcal{Z}[j;t)\,,\\
	\frac{\delta^{2}\mathcal{Z}[j;t)}{\delta j_{r}(\tau)\,\delta j_{r}(\tau')}&=-\mathfrak{J}_{q}(\tau)\mathfrak{J}_{q}(\tau')\,\mathcal{Z}[j;t)\,,
\end{align}
and
\begin{align}
	-i\,\frac{\delta\,\Xi[j;\tau')}{\delta j_{q}(\tau)}&=\frac{1}{2}\biggl[\sigma^{2}D_{1}(\tau)D_{1}(\tau')+\frac{1}{m^{2}\sigma^{2}}\,D_{2}(\tau)D_{2}(\tau')\biggr]\\
	&\qquad\qquad\qquad\qquad\qquad\qquad+\frac{e^{2}}{m^{2}}\int_{0}^{\tau}\!ds'\!\int_{0}^{\tau'}\!ds''\;D_{2}(\tau-s')G_{H,\,\beta}^{(\phi)}(s'-s'')D_{2}(\tau'-s'')\,,\notag\\
	\frac{\delta\,\mathfrak{J}_{q}(\tau')}{\delta j_{q}(\tau)}&=\frac{1}{m}\,D_{2}(\tau-\tau')\,,
\end{align}
with $\mathfrak{J}_{q}(s)$, $\Xi[j;\tau)$ defined by
\begin{align}
	\mathfrak{J}_{q}(s)&=\frac{1}{m}\int_{0}^{t}\!ds''\;D_{2}(s''-s)j_{q}(s'')\,,\\
	\Xi[j;\tau)&=\frac{i}{2}\int_{0}^{t}\!ds'\;\Bigl[\sigma^{2}D_{1}(\tau)D_{1}(s')+\frac{1}{m^{2}\sigma^{2}}\,D_{2}(\tau)D_{2}(s')\Bigr]j_{q}(s')\biggr.\\
	&\qquad\qquad\qquad\qquad+\biggl.\frac{1}{m}\int_{0}^{t}\!ds'\;D_{2}(\tau-s')\,j_{r}(s')+i\,\frac{e^{2}}{m}\int_{0}^{t}\!ds\!\int_{0}^{t}\!ds'\;D_{2}(\tau-s)\,G_{H,\,\beta}^{(\phi)}(s-s')\,\mathfrak{J}_{q}(s')\,.\notag
\end{align}
Thus  in the limits $j_{r}$, $j_{q}\to 0$, we find that
\begin{align}\label{E:cdwoewrkgm}
	\langle\hat{\phi}(\mathbf{z},\tau)\hat{\chi}(\tau')\rangle&=\frac{e}{m}\int_{0}^{\tau'}\!ds\;D_{2}(\tau'-s)G_{H,\,\beta}^{(\phi)}(s,\tau)\\
	&\qquad\qquad+e\int_{0}^{\tau}\!ds\;G_{R,0}^{(\phi)}(\tau-s)\biggl\{\biggl[\frac{\sigma^{2}}{2}\,D_{1}(s)D_{1}(\tau')+\frac{1}{2m^{2}\sigma^{2}}\,D_{2}(s)D_{2}(\tau')\biggr]\biggr.\notag\\
	&\qquad\qquad\qquad+\biggl.\frac{e^{2}}{m^{2}}\int_{0}^{s}\!ds'\!\int_{0}^{\tau'}\!ds''\;D_{2}(s-s')G_{H,\,\beta}^{(\phi)}(s'-s'')D_{2}(\tau'-s'')\biggr\}\notag\\
	&\qquad\qquad\qquad\qquad-i\,\frac{e}{2}\int_{0}^{\tau}\!ds\;G_{R,0}^{(\phi)}(\tau-s)D_{2}(s-\tau')-i\,\frac{e}{2}\int_{0}^{\tau'}\!ds\;D_{2}(\tau'-s)G_{R,0}^{(\phi)}(s-\tau)\,.\notag
\end{align}
Note that we pick $\phi_{+}$ and $\chi_{+}$ in \eqref{E:gbretuer}, so Eq.~\eqref{E:cdwoewrkgm} is a form of time-ordered two-point function, instead of the Pauli-Jordan type of two-point function. Thus its real part gives the corresponding expectation value of (one half of) the anti-commutator, or the Hadamard-like function,
\begin{align}
	\frac{1}{2}\,\langle\bigl\{\hat{\phi}(\mathbf{z},\tau),\,\hat{\chi}(\tau')\bigr\}\rangle&=\frac{e}{m}\int_{0}^{\tau'}\!ds\;D_{2}(\tau'-s)G_{H,\,\beta}^{(\phi)}(s,\tau)\\
	&\qquad\qquad+e\int_{0}^{\tau}\!ds\;G_{R,0}^{(\phi)}(\tau-s)\biggl\{\biggl[\frac{\sigma^{2}}{2}\,D_{1}(s)D_{1}(\tau')+\frac{1}{2m^{2}\sigma^{2}}\,D_{2}(s)D_{2}(\tau')\biggr]\biggr.\notag\\
	&\qquad\qquad\qquad\qquad\qquad+\biggl.\frac{e^{2}}{m^{2}}\int_{0}^{s}\!ds'\!\int_{0}^{\tau'}\!ds''\;D_{2}(s-s')G_{H,\,\beta}^{(\phi)}(s'-s'')D_{2}(\tau'-s'')\biggr\}\,.\notag
\end{align}
This can also be written in a compact and elegant form,
\begin{align}\label{E:xxxcwrjnd}
	\frac{1}{2}\,\langle\bigl\{\hat{\phi}(\mathbf{z},\tau),\,\hat{\chi}(\tau')\bigr\}\rangle&=e\int_{0}^{\tau'}\!ds\;G_{R}^{(\chi)}(\tau'-s)G_{H,\,\beta}^{(\phi)}(s,\tau)+e\int_{0}^{\tau}\!ds\;G_{R,0}^{(\phi)}(\tau-s)G_{H}^{(\chi)}(s,\tau')\,,
\end{align}
in which we have added some superscript to distinguish the Green's functions of different subsystems. According to the previous discussions, we in fact need the component $\langle\bigl\{\hat{\phi}_{h}(\mathbf{z},\tau),\,\hat{\chi}(\tau')\bigr\}\rangle$, which is then given by
\begin{align}
	\frac{1}{2}\,\langle\bigl\{\hat{\phi}_{h}(\mathbf{z},\tau),\,\hat{\chi}(\tau')\bigr\}\rangle&=e\int_{0}^{\tau'}\!ds\;G_{R}^{(\chi)}(\tau'-s)G_{H,\,\beta}^{(\phi)}(s,\tau)\,.
\end{align}
Taking the coincident limit of its $\tau'$ derivative gives the power delivered by the free field,
\begin{equation}
	P_{\xi}(\tau)=e^{2}\int_{0}^{\tau}\!ds\;\dot{G}_{R}^{(\chi)}(\tau-s)G_{H,\,\beta}^{(\phi)}(s,\tau)\,.
\end{equation}
In the case of a linear oscillator,  there is a more direct way to compute this power. From the quantum Langevin equation 
\begin{equation}\label{E:dbieusfs}
	\ddot{\hat{\chi}}(t)+2\gamma\,\dot{\hat{\chi}}(t)+\omega^{2}_{\textsc{r}}\,\hat{\chi}(t)=\frac{e}{m}\,\hat{\phi}_{h}(\mathbf{z},t)\,,
\end{equation}
where $\omega_{\textsc{r}}$ is the renormalized frequency due to interaction with the field, and $\gamma=e^{2}/8\pi m$ is the damping constant, we have its solution
\begin{equation}
	\hat{\chi}(\tau)=D_{1}(\tau)\,\hat{\chi}(0)+D_{2}(\tau)\,\dot{\hat{\chi}}(0)+\frac{e}{m}\int^{\tau}_{0}\!ds\;D_{2}(\tau-s)\,\hat{\phi}_{h}(\mathbf{z},s)\,,
\end{equation}
in which $D_{1,2}(\tau)$ is a special set of homogeneous solutions of \eqref{E:dbieusfs}, satisfying $D_{1}(0)=1$, $\dot{D}_{1}(0)=0$ and $D_{2}(0)=0$, $\dot{D}_{2}(0)=1$. The power is then  given by
\begin{align}
	P_{\xi}(\tau)=\frac{e}{2}\,\langle\bigl\{\hat{\phi}_{h}(\mathbf{z},\tau),\,\dot{\hat{\chi}}(\tau)\bigr\}\rangle&=\frac{e^{2}}{m}\int^{\tau}_{0}\!ds\;\dot{D}_{2}(\tau-s)\,\frac{1}{2}\,\langle\bigl\{\hat{\phi}_{h}(\mathbf{z},s),\,\hat{\phi}(\mathbf{z},\tau)\bigr\}\rangle\notag\\
	&=e^{2}\int^{\tau}_{0}\!ds\;\dot{G}_{R}^{(\chi)}(\tau-s)\,G_{H,\,\beta}^{(\phi)}(s,\tau)\,.
\end{align}
Eq.~\eqref{E:xxxcwrjnd} can also be derived in a similar fashion. The field equation
\begin{equation} 
	\bigl(\partial_{t}^{2}-\nabla^{2}_{\mathbf{x}}\bigr)\hat{\phi}(\mathbf{x},t)=e\,\hat{\chi}(t)\,\delta^{3}(\mathbf{x}-\mathbf{z})\,,
\end{equation}
admits a solution $\displaystyle\hat{\phi}(\mathbf{z},\tau)=\hat{\phi}_{h}(\mathbf{z},\tau)+e\int_{0}^{\tau}\!ds\;G_{R,0}^{(\phi)}(\mathbf{z},\tau;\mathbf{z},s)\hat{\chi}(s)$ whereby we find
\begin{align}
	\frac{1}{2}\,\langle\bigl\{\hat{\phi}(\mathbf{z},\tau),\,\hat{\chi}(\tau')\bigr\}\rangle&=\frac{1}{2}\,\langle\bigl\{\hat{\phi}_{h}(\mathbf{z},\tau),\,\hat{\chi}(\tau')\bigr\}\rangle+e\int_{0}^{\tau}\!ds\;G_{R,0}^{(\phi)}(\mathbf{z},\tau;\mathbf{z},s)\,\frac{1}{2}\,\langle\bigl\{\hat{\chi}(s),\,\hat{\chi}(\tau')\bigr\}\rangle\notag\\
	&=e\int_{0}^{\tau'}\!ds\;G_{R}^{(\chi)}(\tau'-s)G_{H,\,\beta}^{(\phi)}(s,\tau)+e\int_{0}^{\tau}\!ds\;G_{R,0}^{(\phi)}(\tau-s)G_{H}^{(\chi)}(s,\tau')\,.
\end{align}
We obtain \eqref{E:xxxcwrjnd} again, showing that the functional method indeed recovers the results by the canonical operator approach. This will be particularly beneficial when we deal with the nonlinear case, in which the operator ordering issue may loom over in the canonical operator approach.

The $P_{\xi}$ due to the nonlinear potential discussed in Sec.~\ref{S:fieryga} can then be obtained in a similar manner.

\end{document}